\documentclass[acus]{JAC2000}

\usepackage{graphicx}

\begin{document}
\title{Results of an RF Pulsed Heating Experiment at SLAC\thanks{Work supported by Department of Energy Contract DE-AC03-76SF00515}}

\author{D. P. Pritzkau and R. H. Siemann, Stanford Linear Accelerator Center, Stanford, CA, 94309, USA}

\maketitle

\begin{abstract} 
Results are reported from an experiment on RF pulsed heating of copper
at SLAC.  Damage in the form of cracks may be induced on the surface
after the application of many pulses of RF.  The experiment consists
of two circularly cylindrical cavities operated in the TE$_{011}$ mode
at a resonant frequency of \mbox{11.424 GHz}.  Each cavity received
\mbox{8.5 MW}, \mbox{1.2 $\mu$s} pulses at \mbox{60 Hz} corresponding
to a calculated temperature rise of \mbox{120 K} on the copper
surface.  After \mbox{5.5$\times$10$^7$} pulses, the experiment was
stopped and the copper surfaces were examined.  Damage is present on
the area of the surface where the maxiumum heating occurred.
\end{abstract}

\section{INTRODUCTION}
RF pulsed heating is a process by which a metal is heated from
magnetic fields on its surface due to high-power pulsed RF.  Since the
heating occurs over a short time, the inertia of the material prevents
expansion and thermal stresses are induced.  If these stresses are
larger than the elastic limit, known as the yield strength, then
damage in the form of microcracks on the surface may occur after many
pulses.  This type of damage is known as cyclic fatigue.  For
fully-annealed OFE copper, we expect this damage to occur above
temperature rises of \mbox{40 K}~\cite{Mu80}.  However, this
expectation is based on a static yield strength.  It may be more
appropriate to use a dynamic yield strength to predict the damage
threshold, which may be two to three times higher~\cite{Ko75,Ne97}.

This experiment to study RF pulsed heating on copper consists of
circularly cylindrical cavities operating in the TE$_{011}$ mode at a
resonant frequency of \mbox{11.424 GHz}.  The dimensions of the
cavities were chosen to maximize the heating on the endcaps which are
the test pieces of the experiment and were designed to be removable.
For more details see~\cite{pr97,pr98,pr99}.

\section{PARAMETERS OF EXPERIMENT}

\subsection{Endcap preparation}
Two cavities were connected to a \mbox{50 MW} X-band klystron through
a magic-tee in order to protect the klystron from reflected power.
Four endcaps are tested at one time with this setup.  The endcaps are
approximately \mbox{22 mm} in radius and made of OFE copper.  They are
cut to a \mbox{class-16} finish and then brazed onto stainless-steel
rings so they could be welded onto stainless-steel pistons that are
inserted into the cavities.  Before welding, the endcaps are diamond
\mbox{fly-cut} to a \mbox{mirror-finish} and cleaned for vacuum with a light
chemical etch.  The damage from pulsed heating depends on the material
and on surface preparation, and the chemical etch may have had
significant influence on the results of the experiment~\cite{Da2000}.

\subsection{Cavity parameters}
The unloaded and external cavity Q's were measured with a network
analyzer before and after the experiment.  These values are listed in
Table~\ref{tab:Q}.
\begin{table*}[t]
\begin{center}
\caption{Cavity Q's for TE$_{011}$ mode}
\begin{tabular}{|c|c|c|c|c|c|}
\hline
\textbf{Cavity \#} & \textbf{Theoretical} & 
\textbf{Initial} & \textbf{Final}
& \textbf{Initial} & \textbf{Final} \\
& \textbf{Q$_0$} & \textbf{Measured Q$_0$} & \textbf{Measured Q$_0$} 
& \textbf{Measured Q$\mathrm{_{ext}}$}
& \textbf{Measured Q$\mathrm{_{ext}}$}
\\ \hline
1 & 21890 & 20350 & 14360 & 12315 & 7690 \\
2 & 21890 & 20610 & 16810 & 12220 & 7140 \\
\hline
\end{tabular}
\label{tab:Q}
\end{center}
\end{table*}

The diagnostic TE$_{012}$ mode operating at \mbox{17.8 GHz} was not
used in this experimental run~\cite{pr97,pr98,pr99} because
interference with other modes prevented measurements of dynamic
temperature rise and Q change.  This problem should be resolved for
the next run.

\section{RESULTS}
Each cavity received \mbox{1.2 $\mu$s}, \mbox{8.5 MW} square pulses
from the klystron.  The surface magnetic field at the endcaps is
radial with a $J_1$ variation along the radius.  The maximum field and
temperature rise occur at a radius of \mbox{10.6 mm}.  The calculated
maximum temperature rise on the surface is \mbox{120 K} based on
measurement of the input power.  For details about this calculation
see~\cite{pr97,pr98}.  The experiment was run for 5.5$\times$10$^7$
pulses at this temperature rise.  After completion, final measurements
of the cavity Q's were taken and are listed in Table~\ref{tab:Q}.  The
endcaps were then removed and examined.

RF breakdown occured at the coupling irises and initially limited the
input power to \mbox{8.5 MW}.  Although the cavity eventually
processed, we elected to remain at a temperature rise of \mbox{120 K}
to simplify the accounting of the pulses.  After the end of the
experiment, we discovered the coupling irises were melted on the
inside of the cavity.  Since the holes became effectively larger, the
coupling to the TE$_{011}$mode was increased.  This explains the
reduction in the external Q's.

\begin{figure}[htb]
\centering
\includegraphics*[width=70mm]{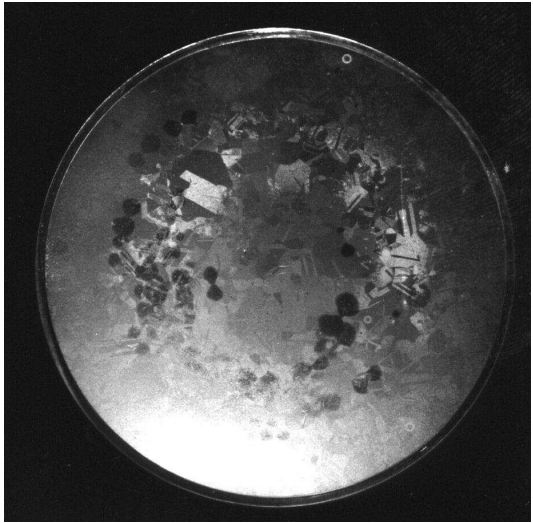}
\caption{One of the copper endcaps.  The other endcaps are similar in
  appearance.}
\label{fig:Macro}
\end{figure}

In Figure~\ref{fig:Macro}, one can see a visible change in the crystal
grains in the region approximately \mbox{10.4 mm} to \mbox{11.0
mm} from the center with an average width of approximately \mbox{4 mm}
to \mbox{5 mm}.  
From GdifdL~\cite{Br2000} simulations, the fields in the region closer
to the coupling iris is 5\% higher than the fields diametrically
opposite of this region.  This simulation would then explain why there
is more damage in the region closer to the aperture, since the region
farther away has a temperature rise approximately 10\% lower.

Several other effects that are not critical to the central goal of
measuring damage from pulsed heating were also observed.  Initial
examinations have shown that copper sputtered onto the endcaps from
the coupling irises.  The density of the copper globules increases
closer to the plane of the coupling irises.  One may also see dark
circles along the area of maximum magnetic field.  Since they seem to
occur in pairs and the magnetic field in this area is radial, we
believe these patches are due to multipacting.  However, it is not yet
understood why this occurs since the electric field in this region is
zero.

\begin{figure}[htb]
\centering
\includegraphics*[width=55mm]{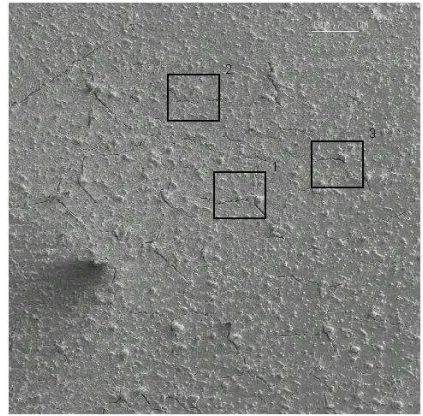}
\caption{Region (\mbox{0.9 mm} $\times$ \mbox{0.9 mm}) where maximum
  damage is expected.  The scale in the upper right-hand corner is
  \mbox{100.7 $\mu$m}.  Figure~\ref{fig:HighMag} has higher
  magnification images of the indicated regions.}
\label{fig:MaxHeatingGrand}
\end{figure}

\begin{figure}[htb]
\centering
\includegraphics[width=55mm]{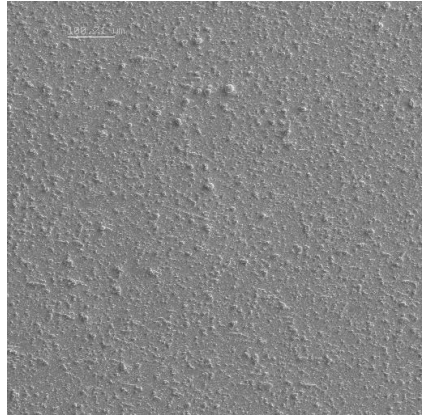}
\caption{Center of endcap.  The scale in the upper left-hand corner is
  \mbox{100.7 $\mu$m}.}
\label{fig:Center}
\end{figure}

The endcaps were examined with a scanning electron microscope using
secondary emission and a \mbox{5 keV} electron beam.
Figure~\ref{fig:MaxHeatingGrand} shows a region where maximum damage
is expected.
Note that this endcap pictured is different from the one shown in
Figure~\ref{fig:Macro}; however, all four endcaps are similar in
appearance.  The little bumps along the surface are the copper
globules sputtered from breakdown at the coupling iris.  More
importantly, one can easily see the numerous cracks that have occured
on the surface.  For comparison, a region at the center where the
temperature rise is close to zero is shown in Figure~\ref{fig:Center}.
This picture was taken at the same magnification.  There are no cracks
in this region.  The copper globules are also noticeable here.
Because of the sputtering, we cannot isolate the unloaded Q
degradation shown in Table~\ref{tab:Q} due to sputtering from that due
to the presence of the cracks.

\begin{figure}[htb]
  \centering
  \includegraphics[width=60mm]{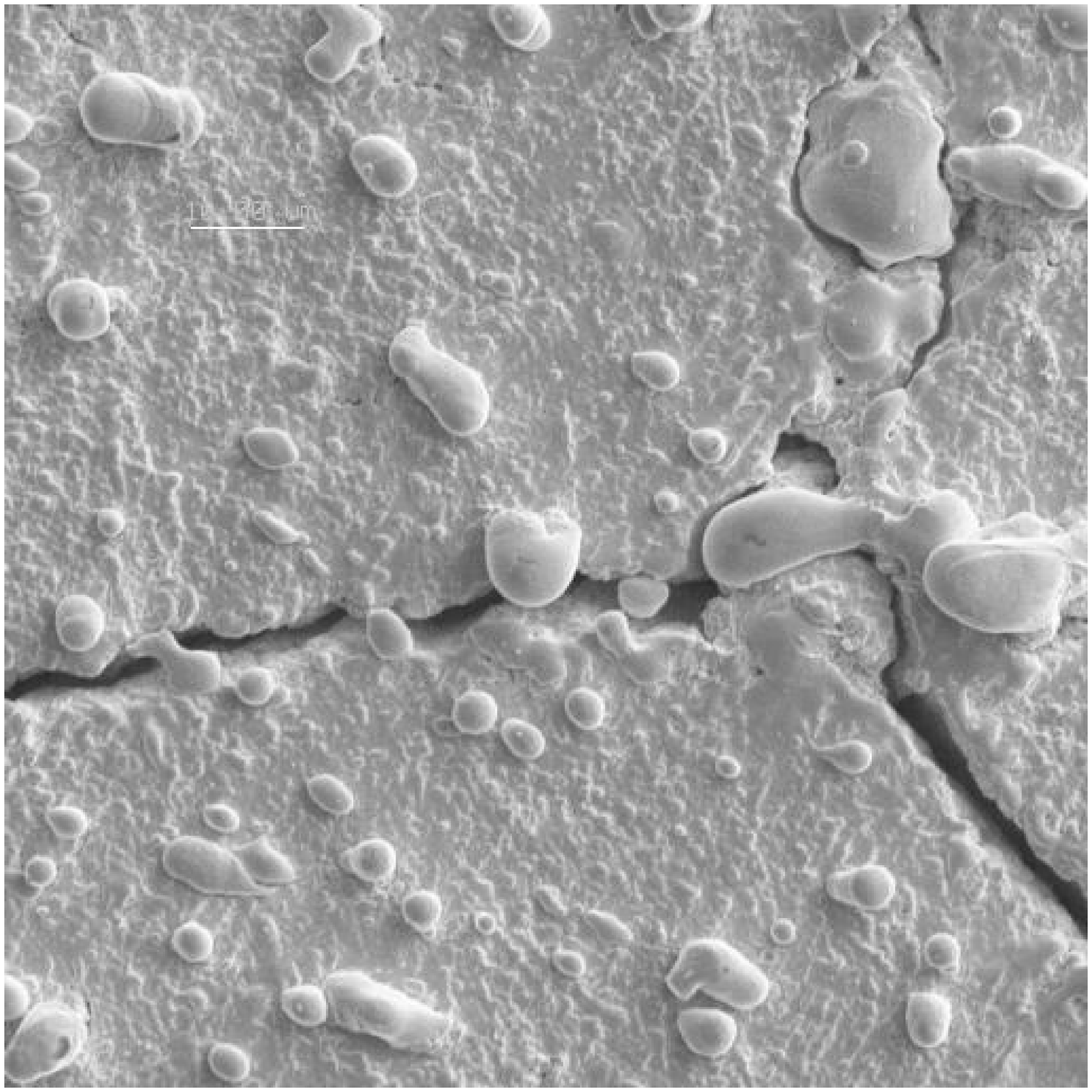}
  \includegraphics[width=60mm]{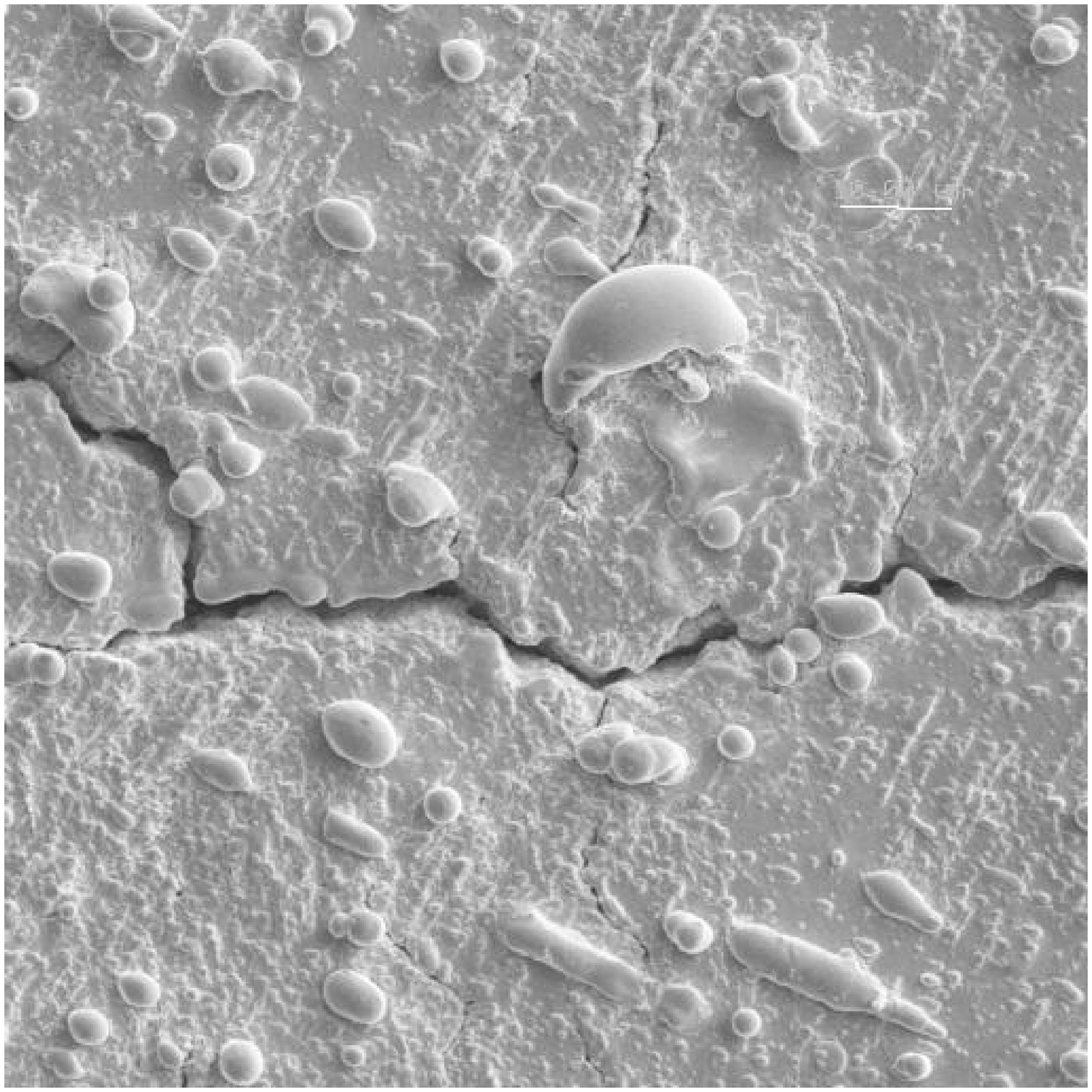}
  \includegraphics[width=60mm]{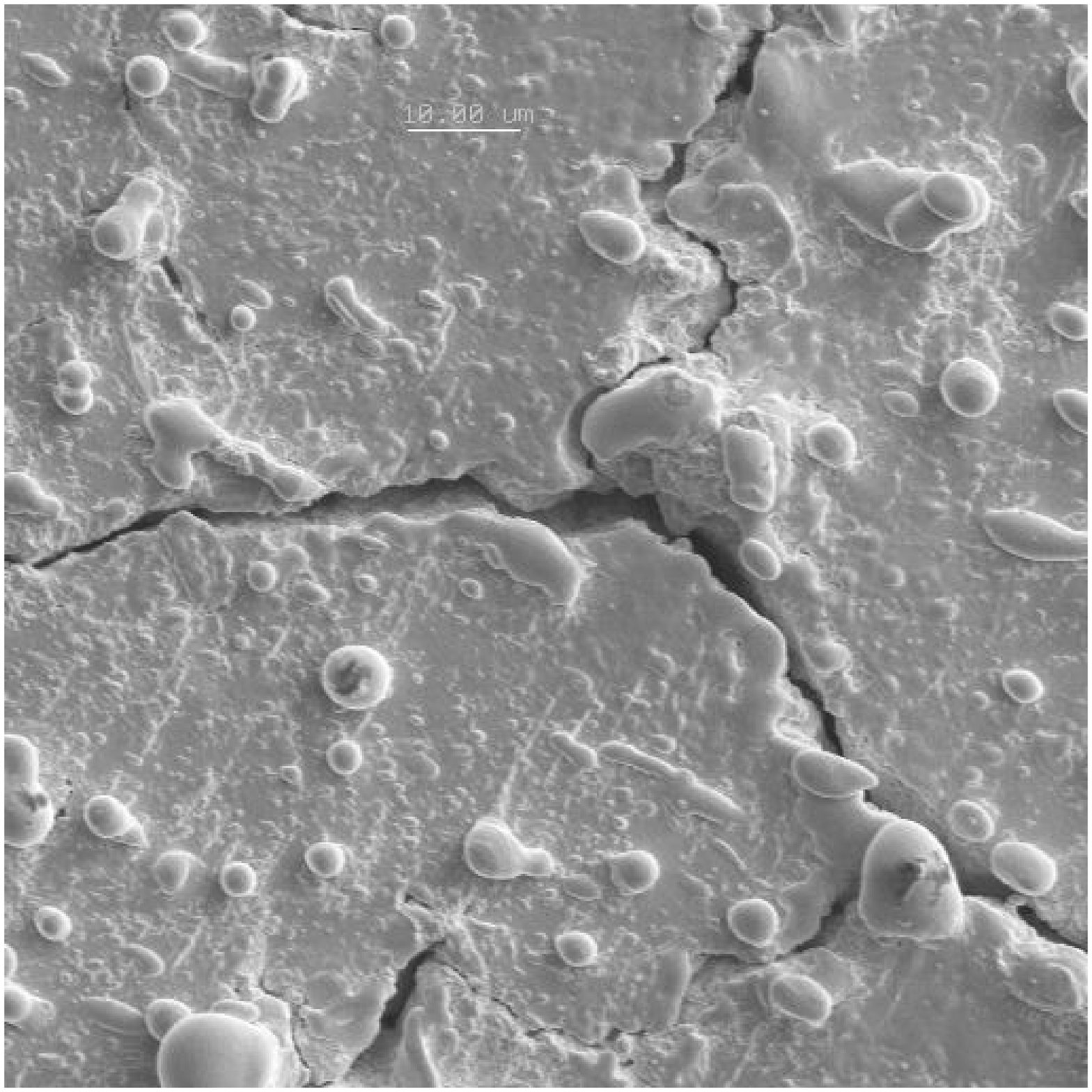}
  \caption{Regions 1--3 from Figure~\ref{fig:MaxHeatingGrand}.  The
    scale shown in each image is \mbox{10.0 $\mu$m}.}
  \label{fig:HighMag}
\end{figure}

Each endcap was scanned along a random diameter.  The scans show that
cracks are only evident in the region of visible grains in
Figure~\ref{fig:Macro} where maximum damage is expected.  Higher
magnification images of the regions of cracks were taken.  For the
endcap shown in Figure~\ref{fig:MaxHeatingGrand}, the average location
of all higher-magnification images of cracks is approximately
\mbox{10.4 mm} from the center with a standard deviation of \mbox{1.3
mm}.  Close-up shots of a few of these cracks are shown in
Figure~\ref{fig:HighMag}.

\begin{figure}[htb]
\centering
\includegraphics*[width=60mm]{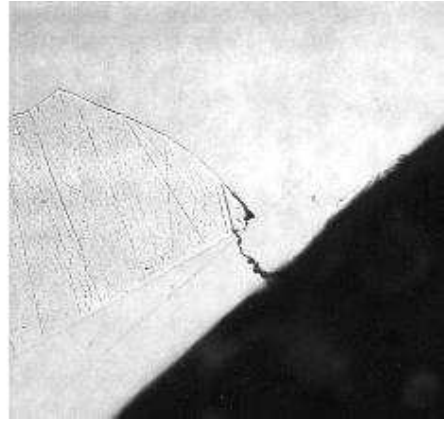}
\caption{A cross-section of the endcap shown in
  Figure~\ref{fig:Macro} showing an internal crack in the
  region of maximum temperature rise.  The crack extends \mbox{24
  $\mu$m} into the surface before touching the grain boundary.}
\label{fig:CrossSectionCrack}
\end{figure}

An internal cross-section of the endcap shown in
\mbox{Figure~\ref{fig:Macro}} was completed.  Not as many cracks were
evident in the cross-section, but they did occur in the region of
maximum temperature rise.  One interesting crack is shown in
\mbox{Figure~\ref{fig:CrossSectionCrack}}.
It begins at the surface and propagates down to an internal
grain boundary.  Afterwards, the crack propagates along the grain
boundary.

\section{CONCLUSION}
We have demonstrated that RF pulsed heating with a temperature rise of
\mbox{120 K} poses a danger to OFE copper structures.  The evidence is
the surface modification and damage shown in
\mbox{Figures~\ref{fig:Macro}, \ref{fig:MaxHeatingGrand},
\ref{fig:HighMag} and \ref{fig:CrossSectionCrack}}.  We have measured
degradation of the cavity Q$_0$ also.  Because of copper sputtered
from breakdown at the coupling iris, it is not conclusive that this
degradation was caused by the surface damage.

We are preparing one more experiment with OFE copper.  The principle
improvement will be use of the diagnostic TE$_{012}$ mode to measure
temperature rise and change of cavity Q during the RF pulse.  This
will also allow measurement of the evolution of Q$_0$ during the
course of the experiment.

RF pulsed heating is a limit to accelerator performance, and
additional experiments need to be done for different temperature
rises, pulse counts, materials and surface preparation.

\section{ACKNOWLEDGEMENTS}
The authors would like to thank Gordon Bowden for his help with
the design of the test cavities.  The authors would also like to thank
Al Menegat for his assistance with running the experiment.


\begin{thebibliography}{9}

\bibitem{Mu80}
H.M. Musal, Jr., ``Thermomechanical Stress Degradation of Metal Mirror
Surfaces Under Pulsed Laser Irradiation'', Laser Induced Damage in
Optical Materials 1979, pp. 159--173.

\bibitem{Ko75}
V.F. Kovalenko, \emph{Physics of Heat Transfer and Electrovacuum
Devices}, ch.7, Moscow: Sovetskoe Radio, 1975.  In Russian.

\bibitem{Ne97}
O.A. Nezhevenko, ``On the Limitations of Accelerating Gradient in
Linear Colliders Due to the Pulse Heating'', PAC'97, Vancouver,
Canada, May 1997.

\bibitem{pr97}
D.P. Pritzkau, et al., ``Experimental Study of Pulsed Heating of
Electromagnetic Cabities'', PAC'97, Vancouver, Canada, May 1997.

\bibitem{pr98}
D.P. Pritzkau, et al., ``Possible High Power Limitations From RF
Pulsed Heating'', 4th RF Workshop (RF98), Watsonville, California,
October 1998.  SLAC-PUB-8013.

\bibitem{pr99}
D.P. Pritzkau, et al., ``Experimental Design to Study RF Pulsed
Heating'', PAC'99, New York, March 1999.

\bibitem{Da2000}
Prof. Reinhold Dauskardt, private communication, 2000.

\bibitem{Br2000}
Warner Bruns, \emph{GdfidL v1.2}, 2000.

\end{thebibliography}
\end{document}